\documentclass[aps,prx,reprint,showpacs,twocolumn,notitlepage]{revtex4-1}
\usepackage[T1]{fontenc}
\usepackage[pdftex]{graphicx}  
\usepackage{graphicx, xcolor}
\usepackage{dcolumn}
\usepackage{bm}
\usepackage{amsmath,amsthm,amssymb}
\usepackage{hyperref}
\usepackage{xcolor}
\hypersetup{colorlinks,bookmarksopen,bookmarksnumbered,
citecolor=blue,
linkcolor=blue,
pdfstartview=false,
urlcolor=blue}
\usepackage{verbatim}
\usepackage{algorithm}
\usepackage[noend]{algpseudocode}

\begin{document}
\title{Measurement-induced criticality as a data-structure transition}

\author{Xhek Turkeshi}
\affiliation{JEIP, USR 3573 CNRS, Coll\`{e}ge de France, PSL Research University, 11 Place Marcelin Berthelot, 75321 Paris Cedex 05, France}

\begin{abstract}
We employ unsupervised learning tools to identify the dynamical phases and their measurement-induced transitions in quantum systems subject to the combined action of unitary evolution and stochastic local measurements. 
Specifically, we show that the principal component analysis and the intrinsic dimension estimation provide order parameters that directly locate the transition and the critical exponents in the classical encoding data space. Finally, we test our approach on stabilizer circuits as proof of principle, finding robust agreement with previous studies.
\end{abstract}
\date{\today}
\maketitle

\section{Introduction}
The advances in noisy intermediate scale quantum devices~\cite{Preskill2018quantumcomputingin,vijay,koh} have motivated a renewed interest in monitored quantum systems~\cite{wiseman_milburn_2009} -- systems where the unitary dynamics is interspersed by local measurements. 
The resulting non-unitary evolution is described by stochastic quantum trajectories stemming from the intrinsic randomness of the quantum measurement operations, that in the many-body framework leads to measurement-induced transitions between unconventional dynamical phases~\cite{Nahum2020,romain1,lunt1,sierant2,romito2}.
These critical phenomena are controlled by the competition between the entangling power of unitary dynamics, which drives the system toward thermalization, and the disentangling effect of local measurements, that collapse the system wave-function in restricted manifolds of the Hilbert space~\cite{Li2018,Li2019,Skinner2019,Szyniszewski2019,Szyniszewski2020,Fan2020,Biella2020,romito4,chen2}.
In the simplest setup of random quantum circuits, these measurement-induced transitions separate a quantum error correcting phase at low measurement rate from a quantum Zeno phase at high measurement rate~\cite{Choi2019,Bao2019,Gullans2019,Gullans2019B}, as shown by extensive numerical investigations~\cite{Zabalo2019,zabalo1,sierant1,romain6,block1,sharma,Lunt2020B,Turkeshi2020,Ippoliti2020} and analytical arguments~\cite{Nahum2020,Jian2019,Piqueres2020,Jian2020,Lang2020,romain2,romain4,romain5,Ippoliti2020B,grover} on the entanglement properties of the system.

\begin{figure}[h!]
    \includegraphics[width=\columnwidth]{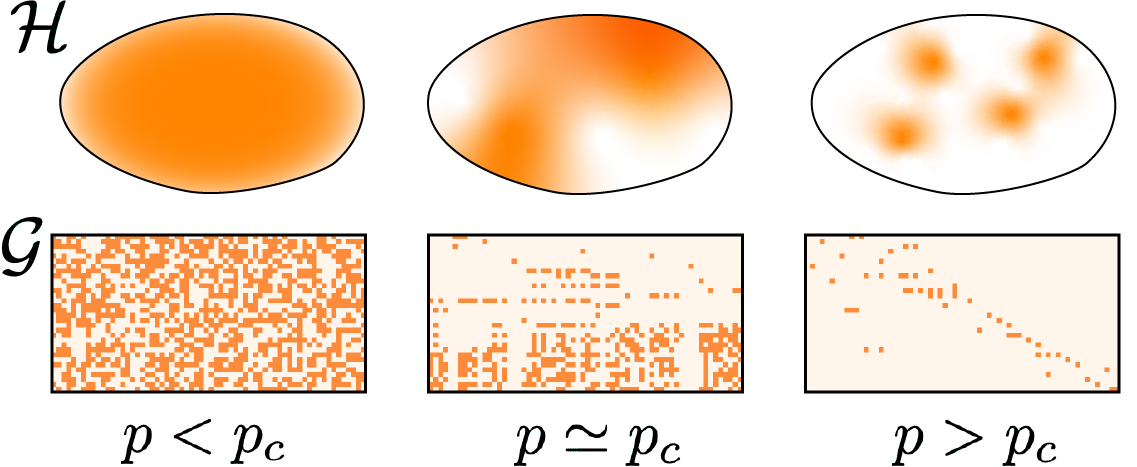}
    \caption{\label{fig:cartoons} Pictorial representation of the phase transition. The Hilbert space $\mathcal{H}$ explored by the quantum trajectories exhibit a structural transition with the measurement rate $p$. This reflects in a geometric transition on the encoding data space $\mathcal{G}$. 
    }
\end{figure}

In this work, we propose an alternative viewpoint by analyzing the classical encoding configurations of the system state and show that the measurement-induced criticality manifests as a geometric transition in the data space (cf. Fig.~\ref{fig:cartoons}). 
To this end, we consider the principal component analysis (PCA) and the intrinsic dimension estimation, which, as unsupervised learning techniques, provide an ideal framework to seek a pattern in unlabelled raw data~\cite{Mehta2019,Mendes2020,Mendes2021}. 
PCA aims to detect the most relevant directions in data space and to compress (project) the data set toward the significant and restricted manifold.
Being a linear method, PCA is particularly effective on linear problems but generally fails when dealing with non-linear structures and complex data space topology~\cite{Geladi2001}. 
On the other hand, the intrinsic dimension estimation extrapolates the effective dimension of the subspace of the data space where the data lie and may be applied to non-linear geometries as well~\cite{Mendes2020,Mendes2021}.

Using stabilizer circuits as a benchmark framework, we argue that the first principal component and the intrinsic dimension are  natural order parameters for the measurement-induced transition, in the same fashion as they are for classical and quantum criticality in equilibrium systems~\cite{Wang2016,Wetzel2017,Scalettar2017,Chng2018,Singh2017,CeWang2017,Scalettar2020,Mendes2020,Melko2018,Lidiak2020,Torlai2020,Martiniani2019,Bagrov2020}. 
(See also Ref.~\cite{Mehta2019,Carleo2019,Carrasquilla2020} for general reviews on machine learning methods in quantum physics).
Furthermore, we find that at the critical point the system develops a minimum intrinsic dimension, which reflects the parametrical simplicity required to describe the system around the transition by virtue of universality. 
Our numerical results perfectly agree with previously reported values of the critical point and the correlation length critical exponent and provide a viable alternative to studying measurement-induced criticality in more general setups.

The remaining of the paper is structured as follows. In Sec.~\ref{sec:methods} we discuss the unsupervised learning methods and how these can be applied to the data space of quantum trajectories. 
In Sec.~\ref{sec:stab} the stabilizer circuits used to benchmark our methods and review how these can be encoded and simulated in polynomial resources, and the relevant results for (1+1)-dimensional systems which we will use for comparison with our analysis. 
Sec.~\ref{sec:results} discuss our main numerical findings on the principal component analysis and the intrinsic dimension estimation. Finally, our concluding remarks and outlooks are presented in Sec.~\ref{sec:conclusion}.

\section{Data space of quantum trajectories}
\label{sec:methods}
In this section, we introduce the principal component analysis and the intrinsic dimension estimation and discuss the effectiveness and limitation when applied to the encoding data set of quantum trajectories. 

For any given quantum trajectory $|\Psi(\alpha,\xi)\rangle$, with $\alpha$ some control parameters, and $\xi$ a registry identifying the trajectory, we define a feature $G_i$ with $i\equiv (\alpha,\xi)$ as the $d$-dimensional classical encoding of a state. The space of the features is denoted by $\mathcal{G}$ and is called data space. 
A few examples are the following, where we consider states defined on a qubit lattice with $R$ sites. (i) The computational basis representation of a quantum state: the feature is the vector whose components are the amplitude with respect to the computational basis, and hence the dimension of the feature is $d=2^R$~\cite{zala}. (ii) A Gaussian quantum state with correlation matrix $C$: the feature is the one-dimensional reshaping of the correlation matrix, with $d\propto R^2$. (iii) The matrix product state (MPS) representation of a quantum state with uniform bond dimension $D$: the feature is the one-dimensional reshaping of the MPS with $d\propto R D^2$. (iv) A stabilizer state: the feature is the one-dimensional reshaping of the tableau representation and $d=R(2R+1)$ (See Sec.~\ref{sec:stab} and  Ref.~\cite{Nielsen2012,Aaronson2004}). 

A data set is a rectangular matrix $G({\tilde{\alpha},\tilde{\xi}})$ of dimension $N\times d$, where each row is a feature $G_i$ and $N$ is the total number of features, which can include different values of $\alpha$ and of $\xi$. We denote $\tilde{\alpha}$ ($\tilde{\xi}$) the common parameters (post-selected trajectories) of the data set. 
Despite from these data sets one can, in principle,  compute the physical properties of the system (e.g. the entanglement entropy and the correlation functions), here we argue that the measurement-induced criticality emerges as a geometric transition in the data space $\mathcal{G}$, i.e. in the $d$-dimensional space of all the features (cf. Fig.~\ref{fig:cartoons}). 

\subsection{Principal component analysis}
\label{subsec:PCA}
Principal component analysis (PCA) is a projective method based on a linear transformation of the data space basis~\cite{Geladi2001,Mehta2019,Wang2016}.
Following Ref.~\cite{Wetzel2017,Scalettar2017}, we consider as data set a collection of $N_\xi$ quantum trajectory snapshots for each of the $N_\alpha$ values of the parameter $\alpha$. (In this case, there are no shared parameters  $\tilde{\alpha}$ or $\tilde{\xi}$ among the features). These $N=N_\alpha N_\xi$ features are identified as vectors in a $d$-dimensional space. The PCA rotates the framework of reference, in such a way that the variance of the data is the largest in the first transformed direction, the second largest in the second direction, \emph{etc.}. 

The method consists of three steps. (i) Define the centered data set $X$, whose elements are $X_{i,j} = G_{i,j}-(1/N)\sum_i G_{i,j}$ and compute the matrix $\Sigma = X^T X/(N-1)$. The centering preprocess guarantees that this is the covariance matrix of the data set, whose elements are the cross-correlations $\Sigma_{i,j}$ among features. (ii) Compute the eigendecomposition $\Sigma = V^T K V$, where $K=\mathrm{diag}(k_1,\dots,k_d)$ is the diagonal matrix of the eigenvalues ordered in descending order, and $V=(v_1,\dots,v_d)$ is the rotation whose columns $v_j$ identify the $j$-th relevant directions. In the new reference frame defined by $V$, the transformed features have no cross-correlations, and the variance of the data along the $j$-th direction is given by $k_j$. (iii) Rotate the original data set to $W=G V$. The vectors $w_j$ along the direction $v_j$ are termed $j$-th principal component.

A normalized and relative weight of the relevance for the principal components is the explained variance ratios $\lambda_j \equiv k_j/(\sum_i k_i)$~\cite{Mehta2019}. By definition $\sum_n \lambda_n=1$, hence $\lambda_n$ represent the percentage of encoded information along the direction $v_n$. 

Interestingly, in many-body physics at equilibrium, the first principal component acts as an order parameter~\cite{Wang2016,Wetzel2017,Scalettar2017,Chng2018,Singh2017,CeWang2017,Scalettar2020,Mendes2020,Melko2018,Lidiak2020,Torlai2020,Martiniani2019,Bagrov2020}.
In the following, we argue that the first principal component plays the role of order parameter also on monitored quantum systems. 

\subsection{Intrinsic dimension}
\label{subsec:id}
The main limitation of the principal component analysis is rooted in the linear nature of the transformation. Hence, when the data space is non-linear and with complex geometry, the PCA needs non-trivial preprocessing (e.g. Kernel methods~\cite{Mehta2019}) to give meaningful information on the system. 

We overcome this limitation by considering the intrinsic dimension estimation~\cite{Goldt2020,Facco2017}, which aims to estimate the effective dimension $I_d(\alpha)$ of the subspace of the data space where the data lie at varying values of the control parameter (e.g. measurement rate) $\alpha$. 
The data sets are given by $G(\alpha)$ with $N$ quantum trajectory snapshots sharing a fixed value of $\alpha$. For monitored quantum systems, we expect that sparse measurements reflect in a large intrinsic dimension, as the system state will explore arbitrary large regions of the Hilbert space (cf. Fig.~\ref{fig:cartoons}). On the other hand, frequent measurements collapse the dynamics to a restricted manifold with a lower intrinsic dimension, as the wave-function will be strongly localized around the measurement dark states~\cite{sierant1}.

We estimate the intrinsic dimension in a density-independent fashion using the two nearest-neighboring technique (2NN)~\cite{Mendes2020,Mendes2021}. For completeness, here we present the general ideas and the limitation of the method and refer to Ref.~\cite{Facco2017} for an in-depth discussion. 
The method relies on the assumption of \emph{locally} uniform data manifolds. Here, the locality is related to the scale at which we look at the data: the larger the data set, the more resolved the distance between points. (Empirically, a finer scale is inversely proportional to the data set size $N^{-1}$.).
We assume a notion of distance in the data space (e.g. the Hamming distance or the Euclidean distance~\cite{Mehta2019}). 
Under these hypotheses, we can locally represent neighboring features as a uniform hypersphere, and using simple geometric arguments we can identify the intrinsic dimension as detailed below. 

For a given feature $G_i$, we compute the first and second nearest-neighboring distances $r_1(G_i)$ and $r_2(G_i)$ in data space, and the ratio $\mu(G_i)=r_2(G_i)/r_1(G_i)$. The hypersphere distribution of neighboring data $G_i$ induce the distribution of the ratios $\mu$ given by
\begin{equation}
\label{eq:id}
    f(\mu) = I_d \mu^{-I_d - 1}.
\end{equation}
From the cumulative distribution $P(\mu)=\int_{0}^\mu d\mu' f(\mu')$ we obtain
\begin{equation}
\label{eq:id2}
    I_d = -\frac{\ln(1-P(\mu))}{\ln\mu}.
\end{equation}
In practice, the cumulative distribution is numerically estimated, and $I_d$ is obtained through a linear fit. 

The 2NN intrinsic dimension estimation is not predictive when the local uniformity of the data set fails. This is the case when the number of features is too small, but, for discrete data sets, also when the number of features is too large. The latter is understood based on the relationship between $N$ and the resolution of the data manifold: When the typical resolution is finer than the typical distance between data points, the discrete structure of a data set emerges and the local uniformity assumption breaks down.  
Thus, the optimal choice for the number of features $N$ lies in a coarse-grain regime, that is, in practice, empirically estimated.

The intrinsic dimension has been studied in many-body physics in Ref.~\cite{Mendes2020,Mendes2021} where it was found to display a local minimum at criticality, which is approached with a critical finite-size collapse. This minimum has an intuitive explanation: at criticality, physics is universal and controlled by a few relevant fields. 
In the following, we show the intrinsic dimension provides a robust order parameter also for monitored quantum systems. 

For self-consistency and completeness, in the next section, we review the monitored quantum system of interest and recall the numerical estimates in the literature which will serve as benchmarks for our analysis.

\section{Stabilizer circuits}
\label{sec:stab}
\begin{figure}[t!]
    \includegraphics[width=\columnwidth]{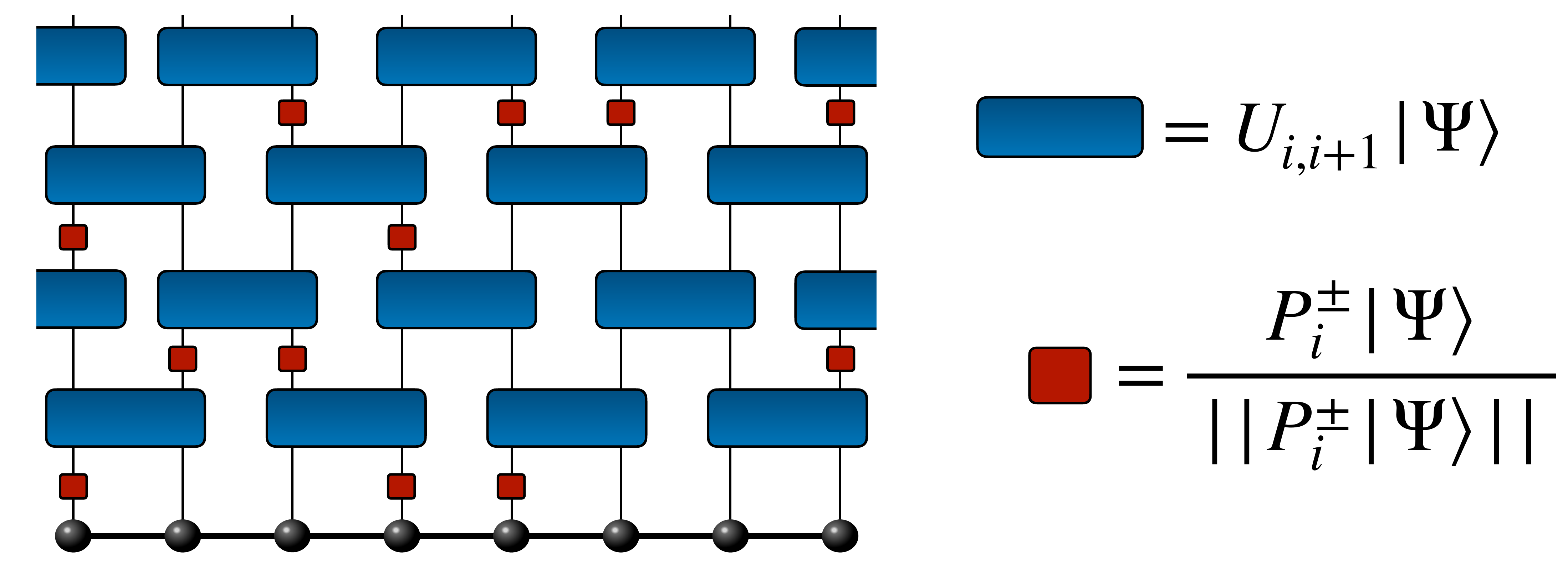}
    \caption{\label{fig:cartoon}Cartoon of the hybrid quantum evolution. The brick-wall unitary is designed to let the qubits propagate correlations, while the measurement gates are randomly peaked with probability $p$. The measurement outcome $\pm1$ determines the collapse operator $P_i^\pm$ through the Born rule. 
    }
\end{figure}

We consider a one-dimensional qubit lattice of size $L$ which evolve through the architecture represented in Fig.~\ref{fig:cartoon}. We assume periodic boundary conditions and $L$ an even number. At each time step, the state evolve according to
\begin{equation}
\label{eq:deftraj}
    |\Psi_{t+1}\rangle = U_t M^{m_t}_t|\Psi_t\rangle,
\end{equation}
where $U_t$, $m_t$ and $M_t^{m_t}$ denote respectively the unitary layer, the measurement outcomes, and the layer of projective measurements at time $t$. We choose  $U_t$ to be a layer of two-body unitary gates given by
\begin{equation}
   U_t = \prod_{i=\mathrm{mod(t,2)}}^{L/2}U_{2 i-1,2i,t}
\end{equation}
with $U_{x,y,t}$ independent random Clifford two-body gates. (A Clifford gate is a unitary gate that map a Pauli string into a \emph{single} Pauli string). The measurement layer is a composition of local measurement operations, which are stochastically picked with probability (measurement rate) $p$. If a local measurement is performed, the resulting qubit is projected onto the measurement result through the Born rule. In summary
\begin{align}
    M^{m_t}_t |\Psi\rangle & = \frac{P^{m_t}_t|\tilde{\Psi}\rangle}{\| P^{m_t}_t|\tilde{\Psi}\rangle\|},\; P^{m_t}_t|\tilde{\Psi}\rangle=\left(\prod_{i=1}^L P_{i,t}^{m_t^i} \right)|\Psi\rangle\nonumber  \\
    P_{i,t}^{m_t^i} &= \displaystyle \begin{cases} \openone & m_t^i=0,\\
        \frac{1\pm Z_i}{2} & m_t^i = \pm 1\end{cases}.
\label{eq:prj}
\end{align}
In a compact fashion, using the time-ordering $\mathcal{T}$ operator, we can write the whole evolution in terms of $K_\mathbf{m} = \mathcal{T}\prod_{t=0}^T (U_t P^{m_t}_t)$ as 
\begin{equation}
    |\Psi_T\rangle = \frac{K_\mathbf{m} |\Psi_0\rangle}{\|K_\mathbf{m} |\Psi_0\rangle\|},
\end{equation}
where $\mathbf{m}$ is a short-hand for the measurement-results and for the unitary gate chosen. The late time regime does not depend on the initial condition, hence without loss of generality we fix the initial state $|\Psi_0\rangle=|0\dots 0\rangle$.

The rate of measurement $p$ controls the dynamical phases of the system~\cite{Li2018}. When the local measurements are suppressed $p\to 0$, the dynamics is governed by the unitary part, which leads the system to explore large manifolds of the Hilbert space at long times. In this regime, measurements are not able to resolve the state of the system, which hence results in a quantum error-correcting phase.
In contrast, frequent measurements $p\to 1$ prevent ergodic behavior as the system is incessantly projected in a reduced manifold (quantum Zeno phase)~\cite{facchi,facchi2}. 

\begin{figure*}[t!]
    \includegraphics[width=\textwidth]{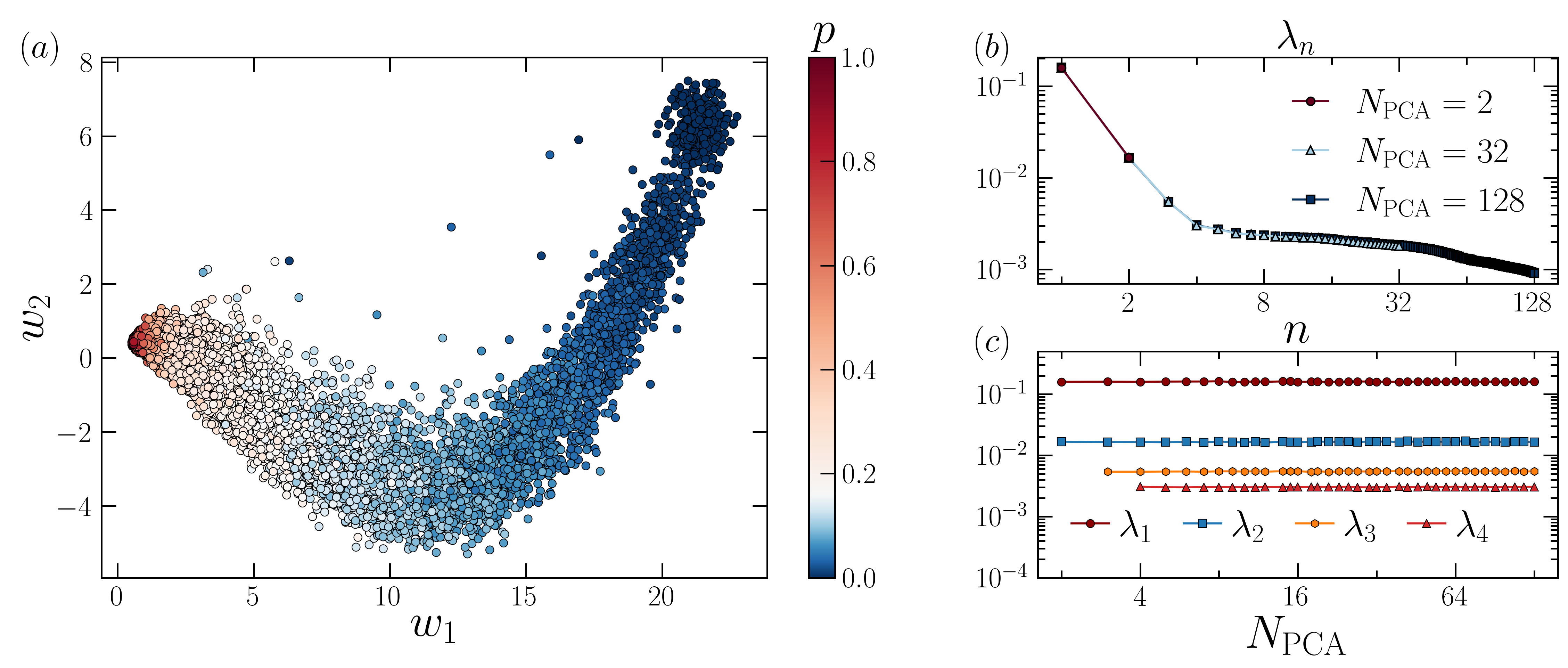}
    \caption{\label{fig:pca1} (a) Results for the principal components $w_1$ and $w_2$ at $L=32$. The data are organized in separate regions for different measurement rates. (b) Explained variance ratios $\lambda_n$ for the most relevant components. (c) The relative relevance of the directions does not change upon increasing the number of components $N_\mathrm{PCA}=2\div 128$.  }
\end{figure*}

With the above specifications, the model is a stabilizer circuit, i.e. a random quantum circuit whose state is a stabilizer at every time step. 
Stabilizer states on $L$ qubits are states for which there exists a subgroup of Pauli strings 
\begin{equation}
    g = e^{i \pi \phi} X_1^{n_1} Z_1^{m_1} X_2^{n_2} Z_2^{m_2}\dots X_L^{n_L} Z_L^{m_L},
\end{equation}
with $\phi,n_j,m_j\in \{0,1\}$ such that $g |\Psi\rangle = |\Psi\rangle$. (We denote $X$, $Y$, $Z$ the Pauli matrices). This group, denoted throughout this paper $Q$, is abelian, and if it is generated by $L$ independent Pauli strings $\hat{g}_j$, it uniquely specifies the system state as
\begin{equation}
\label{eq:deffff}
    |\Psi\rangle\langle \Psi| = \prod_{j=1}^L \left(\frac{1+\hat{g}_i}{2}\right)=\frac{1}{2^L}\sum_{g\in Q} g.
\end{equation}
Since a stabilizer state is encoded in the generating Pauli strings $\hat{g}_i$ (cf. Eq.~\eqref{eq:deffff}), a random Clifford gate $U$ maps a stabilizer state into a stabilizer, fixed by the new stabilizers $U \hat{g}_i U^\dagger$.

In a similar fashion, projective measurements on a Pauli string, map a stabilizer state into a stabilizer state. To see this, consider the measurement on the Pauli string $g_s$. If $[g_s,\hat{g}_j]=0$ for all the generators $\hat{g}_j$ of $Q$, the state of the system is unaffected by the measurement, and the measurement result is deterministic~\footnote{Determining measurement result require the inversion of linear systems in the field $\mathbb{F}_2$.}. 
If this is not the case, there exists a set $\{g_{r_1},\dots,g_{r_l}\}$ that do not commute (but anticommute) with $g_s$. The measurement result is random with probability $1/2$, and the projection onto the measurement result $\pm g_s$ is added to the generators. The commuting generators are left untouched, while the anticommuting set is reduced to $\{g_{r_1}\cdot g_{r_2},g_{r_2}\cdot g_{r_3},\dots,g_{r_l-1}\cdot g_{r_l}\}$ (this certifies that all the generators commute, as it should be). 
The above observations constitute the Gottesman-Knill theorem~\cite{Aaronson2004,Nielsen2012}.

An important consequence is that stabilizer circuits are encoded and simulated in polynomial resources. In particular, a stabilizer state is fixed by the $L\times (2L+1)$ matrix
\begin{equation}
    \hat{G} = \begin{pmatrix} \vec{\phi} &  M_X &  M_Z
    \end{pmatrix}\label{eq:tabbbb}
\end{equation}
where $\phi^j$ is the vector defining the phases, $M_X=[n_i^j]$ is the matrix defining the $X$ operators, and $M_Z=[m^j_i]$ the matrix of $Z$ operators  of the generators $\hat{g}_j$.
In a similar fashion, random Clifford gates and projective measurements represent maps in the $\mathbb{F}_2$ field of the matrix $\hat{G}$. 

We note that the tableau representation is not unique. A particular instance of $\hat{G}$ corresponds to fixing a basis on the stabilizer group $Q$ for the state $|\Psi\rangle$, but any other choice of independent generators $\hat{G}'$ for the stabilizer group $Q$  corresponds to the same state $|\Psi\rangle$. This redundancy is denoted as gauge freedom of the tableau representation. With $|\Psi_0\rangle=|0\dots0\rangle$, we shall fix the gauge fixing the initial tableau $M_Z=\openone$, $M_X=0$  and $\vec{\phi}= 0 $, and update the stabilizer group according to the measurement prescription discussed in this section. However, in discussing physical results, we shall compare our findings with randomized choices of the basis for $Q$. 

The stabilizer circuit in Fig.~\ref{fig:cartoon} exhibits a measurement-induced phase transition at $p_c=0.1599(1)$ with correlation length critical exponent $\nu=1.27(1)$~\cite{Gullans2019B,sierant3}, between a quantum error correcting phase at $p<p_c$ and a quantum Zeno phase at $p>p_c$. 
We shall use this model in the next section to benchmark the methods discussed in Ref.~\ref{sec:methods}.

\begin{figure*}[t!]
    \includegraphics[width=\textwidth]{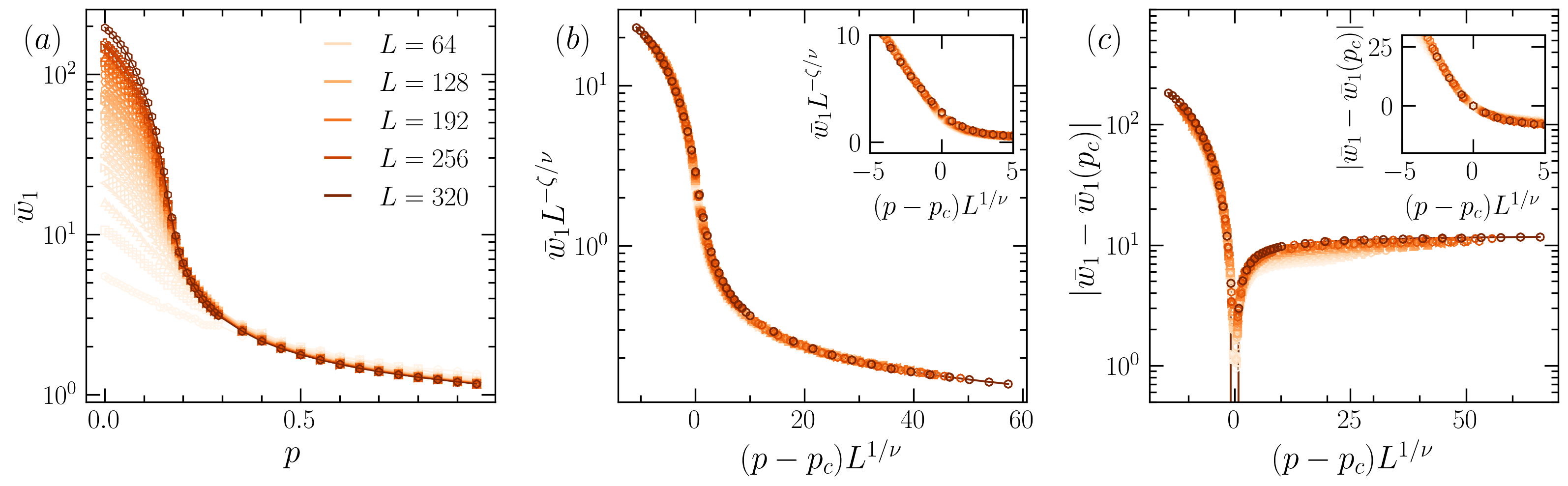}
    \caption{\label{fig:pca2} (a) Quantified principal component $\bar{w}_1$ for different system sizes $L$, and (b,c) its data collapses. The results show the order parameter nature of  $\bar{w}_1$. (b) The estimated $p_c=0.159(4)$, $\nu=1.35(5)$, and $\zeta=0.51(3)$ are in agreement with the results in literature. (c) Also the estimated $p_c=0.165(6)$ and $\nu=1.30(7)$. are in agreement with the results in the literature.  In the insets, we magnify the data collapses close to the critical point.}
\end{figure*}

\section{Numerical benchmarks} 
\label{sec:results}
We implement the stabilizer circuit in Sec.~\ref{sec:stab} using the efficient library \texttt{STIM}~\cite{STIM} based on the algorithm introduced in Aaronson-Gottesman algorithm~\footnote{
The measurement layer described in Sec.~\ref{sec:stab} would require $O(L^3)$ computational resources since, for deterministic measurements, revealing the measurement result $|0\rangle$ or $|1\rangle$ would need inverting a matrix in $\mathbb{F}_2$. 
In Ref.~\cite{Aaronson2004} the authors optimize the measurement layers from $O(L^3)$ to $O(L^2)$ by considering an additional $L\times (2L+1)$ tableau (of destabilizing generator). We refer to Ref.~\cite{Aaronson2004} for a detailed explanation of the algorithm and here mention that these are numerical tools and are not stored as features and are neglected in the learning algorithms. }.
We evolve the state at times $t\ge 8L$, and store the encoding tableau every $\Delta t=L/2$ time-steps. From the $L\times (2L+1)$ tableau representation $\hat{G}_i$ we obtain the feature $G_i$ through reshaping to a  $d=L(2L+1)$ binary vector (cf. Sec.~\ref{sec:methods}).

For any system size $L$, we construct a data set of $N=N_p N_s$ features for the principal component analysis, with $N_p$ the number of values $p\in[0,1]$ considered, and $N_s$ the number of snapshots for each value of $p$~\footnote{We fix $N_p = 43$, with $p\in[0.0,0.01,\dots,0.29]\cup[0.35,0.4,0.45,\dots,0.95]$, and vary $N_s=200,400, 800, 1600$. We present data only for $N_s = 400$, as we find no qualitative behavior on the results.  The system size considered for the PCA range between $L=16\div 320$.}.
For the intrinsic dimension estimation, we have $N_p$ separate data sets each with $N_s$ features obtained at a fixed $p\in[0,1]$. 
Both the PCA and the intrinsic dimension estimation are implemented using the library \texttt{sklearn}~\cite{sklearn}.

\subsection{Principal component analysis}
We begin by discussing the results of the principal component analysis. We truncate the PCA to $N_\mathrm{PCA}$ principal components for efficiency. In fact, from the centered data set $X$ (cf. Sec.~\ref{sec:methods}) we can obtain the principal directions and weight via singular value decomposition, simplifying the computational complexity of the problem. 

As an illustrative example, we present the results of the PCA for $L=32$ in Fig.~\ref{fig:pca1} varying the maximum number of components $N_\mathrm{PCA}$. 
We see that the first principal direction alone captures around $16\%$ of the data set, and within the first $4$ component the cumulative encoding reaches 
$20\%$. (A large portion considered that the dimension of the feature space is $d=L(2L+1)$).
This fact is unaffected by varying the number of directions required by the algorithm, as the explained variance ratios remain qualitatively unchanged. Conversely, $\lambda_n$ distribute into the same curve over the range of considered principal directions $N_\textup{PCA}$. We stress that the data set considered in each case is different, and the small fluctuations are related to the specific realizations. 
Finally, we note the discrete binary nature of the data does not allow for a neat clustering of the data points for $p<p_c$ and $p>p_c$ (for some critical rate $p_c$). The same would occur also considering various kernel methods, and stem from the equivalence between different metrics for discrete binary data, including Euclidean and Hamming distances. This phenomenon should be contrasted with, e.g., Ref.~\cite{long2020unsupervised} where different phases clearly separate through a diffusion map algorithm. 
Finding suitable clustering algorithm for discrete data is an open field of investigation and is left for future investigation.

Although the principal components contain all the relevant information of the data set, it is convenient to extract a meaningful number depending on the value of the measurement rate $p$. 
We consider the quantified principal components, defined as the conditional averages
\begin{equation}
\bar{w}_j = \frac{1}{N_s} \sum_{i(p)} w_{j}(i).
\end{equation}
Here the mean is over the $N_s$ configurations with the same measurement rate $p$. We present the numerical data in  Fig.~\ref{fig:pca2} (a) for various $L$ and $p$, that suggest the presence of a finite size scaling. 

We choose two finite size scaling hypothesis. First, we consider the generic finite-size scaling hypothesis
\begin{equation}
\label{eq:ffss}
\bar{w}_1(p,L) = L^{\zeta/\nu} f_1( (p-p_c) L^{1/\nu}),
\end{equation}
in the spirit of statistical mechanics order parameters. This ansatz is a starting point for models where we do not have \emph{ab-initio} knowledge. 

Furthermore, we also consider an \emph{a fortiori} finite-size scaling hypothesis. It is motivated by the logarithmic corrections present for the entanglement entropy for the measurement-induced criticality of (1+1)D stabilizer circuits~\cite{Li2019}
\begin{equation}
\label{eq:ffss2}
|\bar{w}_1(p,L)-\bar{w}_1(p_c,L)| = \tilde{f}_1( (p-p_c) L^{1/\nu}).
\end{equation}
We neglect the smallest system sizes and consider $L\ge 64$. Performing the finite size scaling with standard techniques~\cite{Zabalo2019}, we find an excellent data collapse for both the hypothesis, as demonstrated in Fig.~\ref{fig:pca2}. For Eq.~\eqref{eq:ffss} our estimate for the critical point and exponents are: $p_c=0.159(4)$, $\nu=1.35(5)$ and $\zeta=0.51(3)$. 
Instead, for Eq.~\eqref{eq:ffss2} we have $p_c=0.165(6)$ and $\nu=1.30(7)$. 
Given our numerical data, we cannot differentiate which scaling is the correct one as their estimates for $p_c$ and $\nu$ are compatible. Nevertheless, the analysis demonstrate that $\bar{w}_1$ is an effective order parameter for the measurement-induced phase transition.

Importantly, $\bar{w}_1$ does not have a straightforward physical interpretation. In general it is a non-local order parameter, as it depends non-trivially on full correlation pattern in the data space. The advantage compared to physically motivated observables (e.g. correlation functions) is that it can be successfully applied also in problems which lack a local order parameter, such as the Berezinskii-Kosterlitz-Thouless transitions or lattice gauge theories~\cite{bkt,wetz}.

Next, we consider the subsequent (less relevant) components, and compute the  quantified principal components $\bar{w}_k$ with $k\ge 2$. We find these exhibits a non-monotonic behavior with the measurement rate $p$, with oscillations appearing in the error-correcting phase ($p<p_c$), while saturating at a $O(1)$ value in the quantum Zeno phase ($p>p_c$) (See Fig.~\ref{fig:pca3}).
These oscillations are due to the choice of gauge-fixing of the tableau representation we have considered in Sec.~\ref{sec:stab}.

To test the gauge dependence of our results, we consider a choice of random generators for the stabilizer group $Q$ fixing the state.
This is obtained through random linear rank-preserving linear combinations of the rows of $\hat{G}_i$ on the field $\mathbb{F}_2$.

As anticipated, the secondary quantified principal component exhibit a qualitative change of behavior at a low-measurement rate, with an $O(L)$ non-monotonic value in the quantum error-correcting phase. At a high measurement rate, the quantified principal components  $\bar{w}_{k\ge 2}$ is $O(L)$  saturate to a constant value. (See $\bar{w}_2$ in Fig.~\ref{fig:pca3} (Right), although similar features are present for the subsequent principal components). 

\begin{figure}[t!]
    \includegraphics[width=\columnwidth]{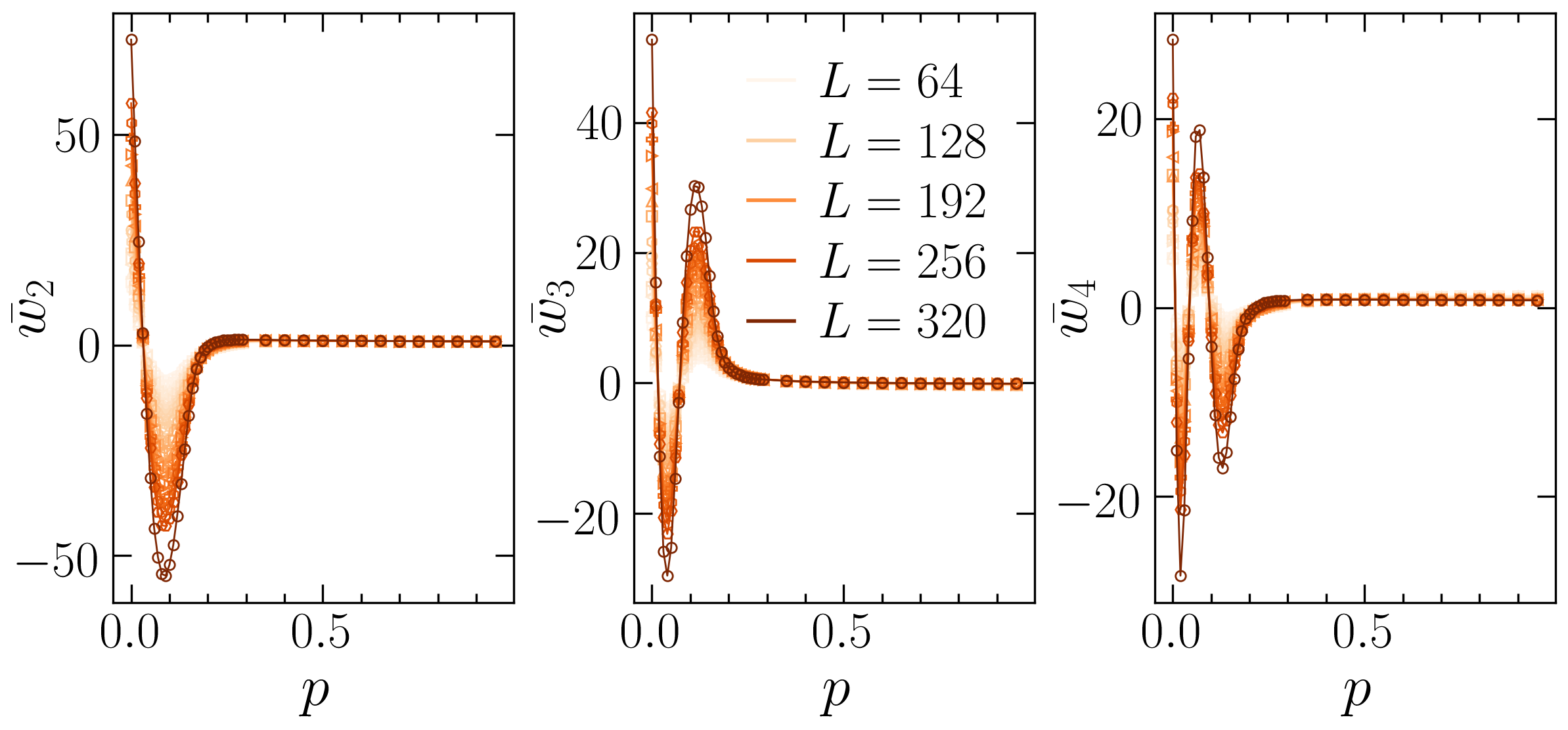}
    \caption{\label{fig:pca3} Secondary quantified principal component for different system sizes $L$. The oscillatory behavior is due to the choice of gauge fixing for the stabilizer tableau representation.  }
\end{figure}

On the other hand, the first quantified principal component exhibit the same qualitative behavior as in Fig.~\ref{fig:pca2} (cf.~Fig.~\ref{fig:pca3} (Left)). Performing the finite size scaling under the hypothesis Eq.~\eqref{eq:ffss}, we find $p_c=0.159(7)$, $\nu=1.35(8)$ and $\zeta=0.52(4)$, in agreement with the estimates in Fig.~\ref{fig:pca3}. 
As a result, the first principal component accesses the universal content of the monitored quantum system within the classical encoding space without prior knowledge or choice of the specific observable.

\begin{figure}[t!]
    \includegraphics[width=\columnwidth]{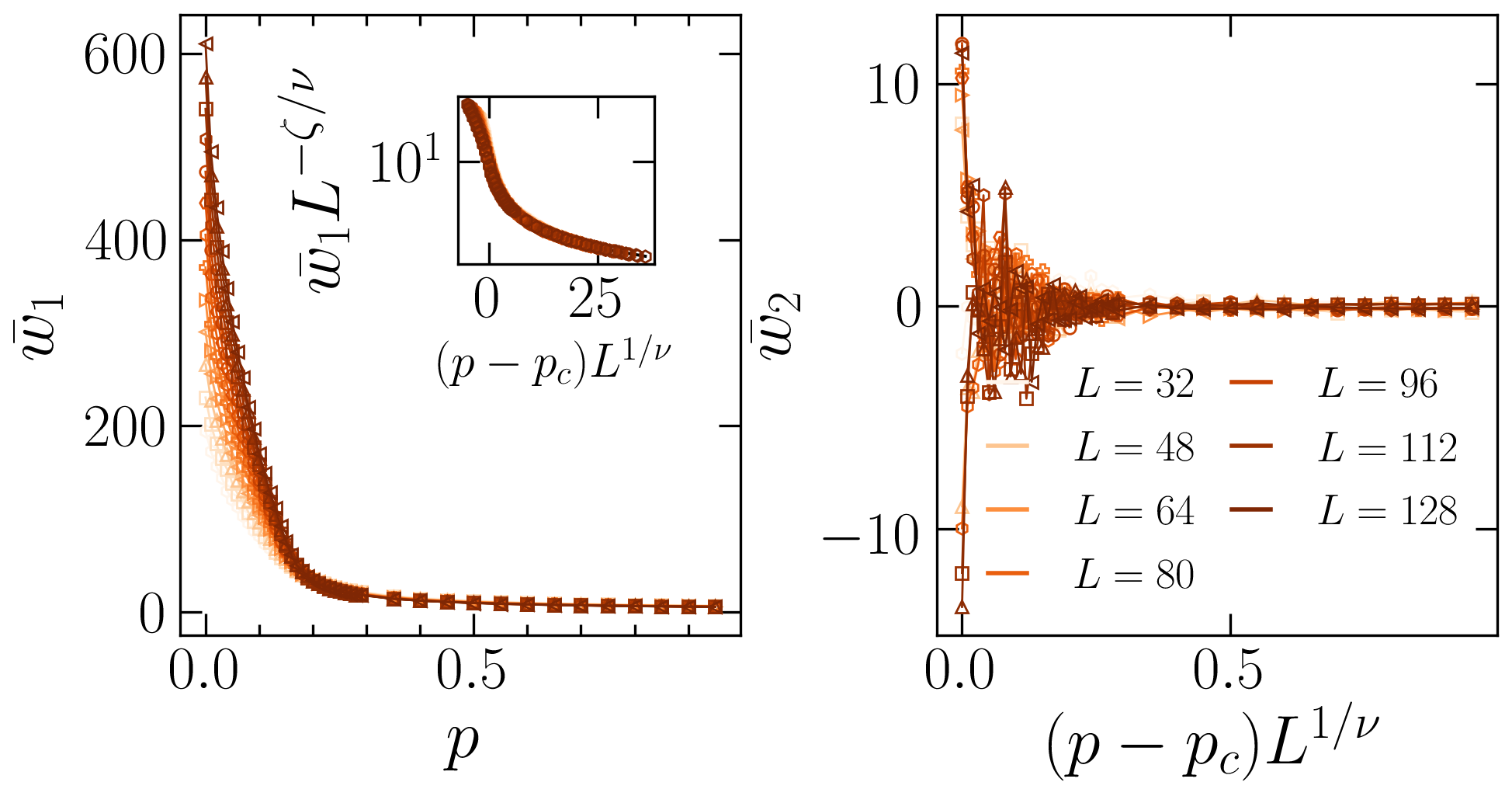}
    \caption{\label{fig:pca3} First (Left) and second (Right) quantified principal component obtained through a random choice of tableau. (Inset) Data collapse  $\bar{w}_1 = L^{\zeta/\nu} f((p-p_c)L^{1/\nu})$ with  $p_c=0.159(7)$, $\nu=1.35(8)$ and $\zeta=0.52(4)$. These results are compatible with the analysis of $\bar{w}_1$ for the specific choice of gauge induced by the algorithm in Sec.~\ref{sec:stab}. }
\end{figure}

\subsection{Intrinsic dimension}
\label{subsec:id}
We next consider how the intrinsic dimension, which is a density-independent quantity applicable to non-linear data spaces, can locate the measurement-induced criticality. 
Given the binary nature of our data points, we consider the Hamming distance defined for two $N$-dimensional vectors $x$ and $y$ as
\begin{equation}
    d(x,y)=\sum_{i=1}^N \delta_{x_i,y_i}.
\end{equation}
With this metric, we perform the 2NN algorithm on the stabilizer configurations. 
For each data point we compute the (next)-nearest neighboring distances ($r_2(G_i)$) $r_1(G_i)$ by computing and sorting $d(G_i,G_j)$ for $j\neq i$.

To obtain a robust estimate of the intrinsic dimension, we collect $N_\mathrm{data}=30$ datasets of $N_s=5000$ for each value of $L$ and $p$ considered, compute the intrinsic dimension over each dataset. Averaging over the $N_\mathrm{data}$ data sets we obtain the final estimate $I_d$~\footnote{We consider $p\in[0.0,0.01,\dots,0.98,0.99]$, and vary the system size in $L=16\div 128$.}.

\begin{figure}[t!]
    \includegraphics[width=\columnwidth]{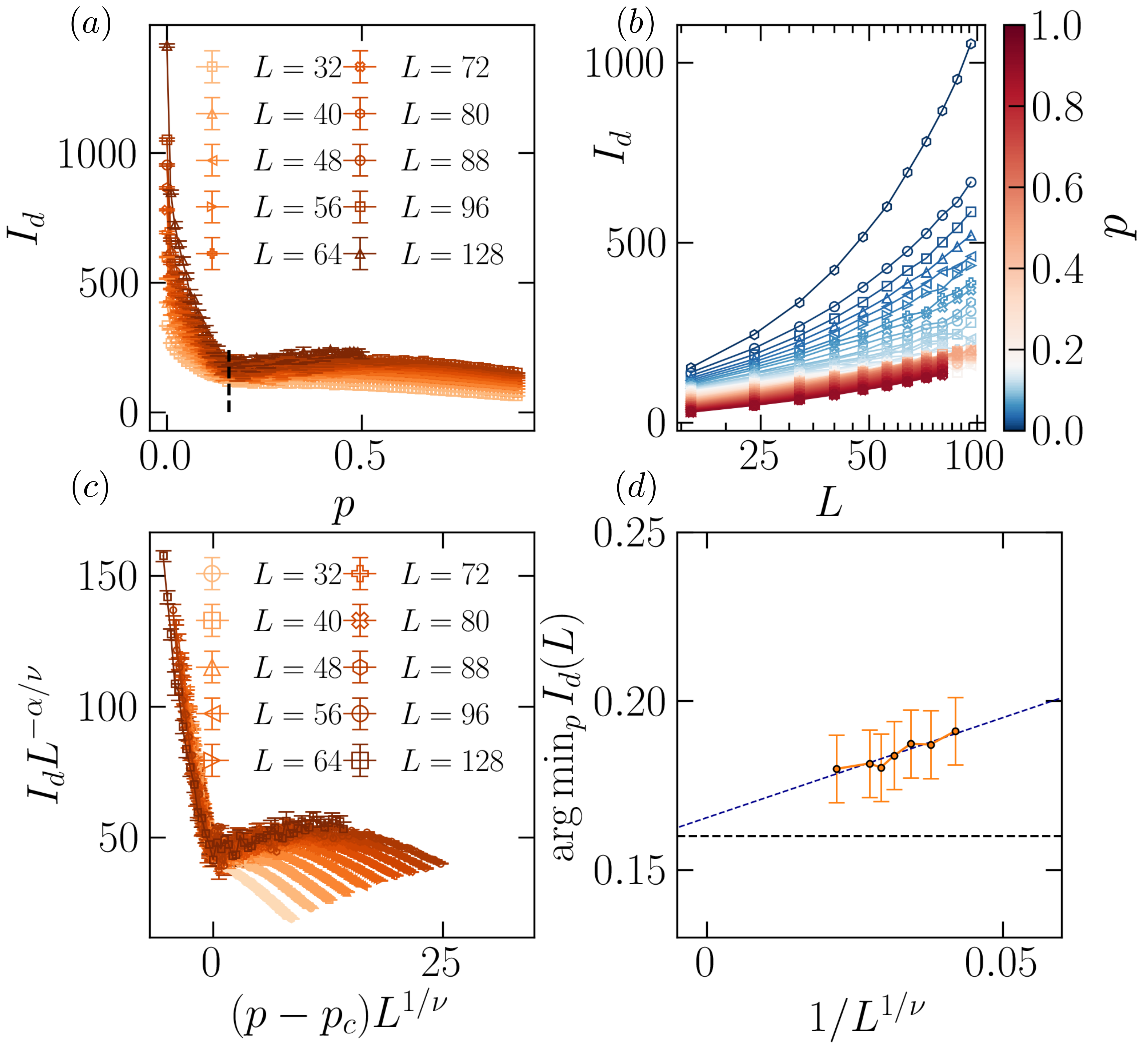}
    \caption{\label{fig:id} (a) Intrinsic dimension for different system sizes $L$. Notice the non-monotonic behavior, with a minimum close to criticality. (b) Scaling of the intrinsic dimension with the system size for various values of the measurement rate. We distinguish a linear region for  $p<p_c$, and a logarithmic one for $p>p_c$. (c) Data collapse after a finite-size scaling analysis. The estimated $\nu=1.3(1)$, $p_c=0.16(2)$, and $\alpha=0.3(1)$, are in agreement with the previous analysis (Fig.~\ref{fig:pca2} and Fig.~\ref{fig:pca3}). (d) Estimation of the critical point through the minimum of the intrinsic dimension.  The points are obtained by fitting a third-order polynomial and locating the minimum. The dashed line is the optimal linear fit in $1/L^{1/\nu}$, where we excluded small system sizes. The intersection $p_c(L\to\infty) = 0.16(2)$ is in agreement with the data collapse. }
\end{figure}

The results are plotted in Fig.~\ref{fig:id}. We find a linear growth of the ID for $p\lesssim 0.16$, while a logarithmic one at $p\gtrsim0.16$. 
The physical interpretation of these results is based on the dimensionality of the Hilbert space. Since the quantum state $\rho$ is obtained by summing over all the stabilizer Pauli strings $Q$ (cf. Eq.~\eqref{eq:deffff}), we have $\mathrm{dim}\mathcal{H} \propto e^{\gamma I_d}$ for some constant $\gamma$. When $I_d$ scales linearly with system size, the Hilbert space explored is exponentially large and the stationary state is a random stabilizer state. Conversely, deep in the Zeno phase, the Hilbert space explored is polynomial in system size. In particular, in the thermodynamic limit, the system is localized in a zero-measure manifold. As remarked before, these considerations are consistent with the results obtained using the entanglement measures~\cite{Li2018}.
Let us stress an important difference: while the entanglement entropy in the Zeno phase saturates, the intrinsic dimension scales logarithmically. This is because the intrinsic dimension is not a measure of entanglement, but include also classical correlations of the encoding data set. 

The intrinsic dimension develops a non-monotonic universal behavior close to criticality. We identify the transition using the data-collapse under the finite-size scaling ansatz 
\begin{equation}
\label{eq:idsca}
     I_d = L^{\alpha/\nu} h((p-p_c) L^{1/\nu}),
\end{equation}
adapting the analysis to values of $p\in [ p^\mathrm{est}_c-\delta p,p^\mathrm{est}_c+\delta p]$ close to the empirically estimated critical point $p_c^\mathrm{est}=0.17$, $\delta p = 0.15$. We obtain $p_c=0.16(2)$, $\nu=1.3(1)$ and $\alpha=0.3(1)$, compatibly with the literature and the PCA analysis in Sec.~\ref{subsec:PCA} (See Fig.~\ref{fig:id}).
In turn, the critical point corresponds to the thermodynamic limit $L\to\infty$ of the local minimum position $p^*(L)\equiv \arg\min_p I_d(L)$. 
At finite size, this minimum is estimated by fitting a cubic function around  $p_c^\mathrm{est}$ and finding the local minimum. 
The phase transition is encoded in a diverging correlation length that, in turns, translates to~\cite{Mendes2020}
\begin{equation}
    p^*(L) - p_c \propto \frac{1}{L^{1/\nu}}.
\end{equation}
Therefore, we expect $p_c =  \lim_{L\to\infty} p^*(L)$, that we obtain by performing a linear fit of $p^*(L)$ against $1/L^{1/\nu}$. Our results are given in Fig.~\ref{fig:id}, and our estimated critical point is $p_c= 0.16(2)$, in agreement with the previous analysis.

This local minimum can be understood by virtue of universality. The critical point is parametrically simpler to describe compared to its vicinity, as irrelevant fields are negligible in the renormalization group sense. However, they play an important role in the off-critical region, which increases the number of parameters close to the transition. 
This picture is \emph{a fortiori} confirmed in the present setup by the presence of a conformal field theory~\cite{Li2019,chen19}, but holds on general ground (i.e. for non-conformal critical points~\cite{Mendes2020,Mendes2021}). 

The critical change of the intrinsic dimension is the hallmark of a geometric transition in the data space. It relates the change in the dimensionality of the Hilbert manifold describing the late time state $|\Psi_T\rangle$ to a change in the classical encoding space.

Lastly, we stress that the intrinsic dimension capture the gauge-independent content of the system. We have performed, but not shown here for readability, the intrinsic dimension estimation for random gauge fixing on the tableau representation $\hat{G}$, and find qualitatively the same results and the same critical value $p_c$ and exponents $\nu$, $\alpha$.

\section{Conclusion and outlooks} 
\label{sec:conclusion}
In this paper, we employed principal component analysis and intrinsic dimension estimation to characterize the measurement-induced phase transition in monitored quantum systems as a geometric transition in the classical encoding data space. 

In full analogy to equilibrium classical physics~\cite{Wang2016,Wetzel2017,Scalettar2017}, the principal component analysis captures the critical behavior and the structural change of the phase for stabilizer circuits. This is exemplified by the first quantified principal component $\bar{w}_1$, which develops a critical finite size scaling around the measurement-induced transition. 

The structural transition is also manifest in the change of the intrinsic dimension, which behaves linearly in the quantum error-correcting phase and logarithmically in the quantum Zeno phase. 
At criticality, the intrinsic dimension develops a local minimum, which reflects the parametrical simplicity of the underlying conformal field theory. 
Overall, our results show full compatibility with the numerical investigation present in the literature, while giving a complementary viewpoint on the nature of the measurement-induced transition. 

The unsupervised character of the considered methods requires no a priori knowledge of the phase space, making them attractive tools in the investigation of monitored quantum systems. 
In this paper, we have focused for simplicity on stabilizer circuits, but the toolbox can be easily adapted to other monitored frameworks, such as Gaussian systems~\cite{buchhold1,buchhold3,buchhold4,buchhold5,buchhold6,romito1,turkeshi1,turkeshi2,linon,legal,turkeshi3,turkeshi4,minato,chen1,chen3,chen4,chen88}, many-body interacting models~\cite{fuji1,tang1,buchhold2,syk1,syk2}, or topological and symmetry-protected topological models~\cite{jans1,buchhold7,romito3,hsieh1,lavasani1,lavasani2}.

Furthermore, principal component analysis can be used to preprocess large data sets in reinforcement and supervised learning methods. We note that such supervised techniques have been recently shown to identify measurement-induced phase transition as a learnability problem~\cite{romain3,lavasani}, and may be suitably adapted to experimental frameworks~\cite{koh,vijay,Noel_2022,melko,sierant2,sierant1}.
Similarly, it would be interesting to extend the unsupervised toolbox for measurement-induced criticality to variational autoencoders~\cite{zala}, which provide an unsupervised neural network method that do not require prior knowledge of the phase diagram.

\begin{acknowledgments}
The author is indebted to M. Dalmonte, R. Fazio, A. Rodriguez, and T. Santos-Mendes for the collaboration on related topics, and their enlightening comments on the manuscript. The author is also grateful to S. Pappalardi, M. Schir\'o, and I. Macocco for discussions. 
The author acknowledges support from the ANR grant "NonEQuMat" (ANR-19-CE47-0001).
\end{acknowledgments}

\bibliography{biblio.bib}
\bibliographystyle{apsrev4-1}

\end{document}